\documentclass[
superscriptaddress,nofootinbib,amsmath,amssymb]{revtex4-2}
\usepackage[colorlinks=true,linkcolor=blue,citecolor=magenta,linktocpage=true]{hyperref}
\usepackage{graphicx}
\usepackage{bm}
\usepackage{hyperref}
\usepackage{mathtools}

\renewcommand{\d}{{\rm d}}
\newcommand{\q}{{\qquad\qquad}}
\newcommand{\f}{\frac}

\newcommand{\mrm}[1]{_{\rm #1}}

\numberwithin{equation}{section}

\makeatletter
\def\l@subsubsection#1#2{}
\makeatother

\begin{document}

\begin{flushright}
CERN-TH-2021-007\vspace*{0.5cm}
\end{flushright}

\title{Hawking radiation by spherically-symmetric static black holes for all spins:\\ I - Teukolsky equations and potentials
\vspace*{0.5cm}}

\author{Alexandre Arbey}
\email{alexandre.arbey@ens-lyon.fr}
\affiliation{Univ Lyon, Univ Claude Bernard Lyon 1,\\ CNRS/IN2P3, IP2I Lyon, UMR 5822, F-69622, Villeurbanne, France}
\affiliation{Theoretical Physics Department, CERN, CH-1211 Geneva 23, Switzerland}
\affiliation{Institut Universitaire de France (IUF), 103 boulevard Saint-Michel, 75005 Paris, France}

\author{Jérémy Auffinger}
\email{j.auffinger@ipnl.in2p3.fr}
\affiliation{Univ Lyon, Univ Claude Bernard Lyon 1,\\ CNRS/IN2P3, IP2I Lyon, UMR 5822, F-69622, Villeurbanne, France}

\author{Marc Geiller}
\email{marc.geiller@ens-lyon.fr}
\affiliation{Univ Lyon, ENS de Lyon, Univ Claude Bernard Lyon 1,\\ CNRS, Laboratoire de Physique, UMR 5672, F-69342 Lyon, France}

\author{Etera R. Livine}
\email{etera.livine@ens-lyon.fr}
\affiliation{Univ Lyon, ENS de Lyon, Univ Claude Bernard Lyon 1,\\ CNRS, Laboratoire de Physique, UMR 5672, F-69342 Lyon, France}

\author{Francesco Sartini}
\email{francesco.sartini@ens-lyon.fr}
\affiliation{Univ Lyon, ENS de Lyon, Univ Claude Bernard Lyon 1,\\ CNRS, Laboratoire de Physique, UMR 5672, F-69342 Lyon, France}


\begin{abstract}
\vspace{0.5cm}
In the context of the dynamics and stability of black holes in modified theories of gravity, we derive the Teukolsky equations for massless fields of all spins in general spherically-symmetric and static metrics. We then compute the short-ranged potentials associated with the radial dynamics of spin 1 and spin 1/2 fields, thereby completing the existing literature on spin 0 and 2. These potentials are crucial for the computation of Hawking radiation and quasi-normal modes emitted by black holes. In addition to the Schwarzschild metric, we apply these results and give the explicit formulas for the radial potentials in the case of charged (Reissner--Nordström) black holes, higher-dimensional black holes, and polymerized black holes arising from loop quantum gravity. These results are in particular relevant and applicable to a large class of regular black hole metrics. The phenomenological applications of these formulas will be the subject of a companion paper.
\end{abstract}

\maketitle

\tableofcontents

\section*{Introduction}
\label{sec:introduction}

Black holes (BHs) are fascinating astrophysical objects. As the ultimate stage of the gravitational collapse of stars, they probe the limits of general relativity and our understanding of high energy and high density physics. With the recent rise of experimental gravitational wave detection, they have become the natural arena to seek and test modified theories of gravity. For this reason, it is essential to analyze all facets of their phenomenology. At the theoretical level, the study of physical properties of black holes sets them at the interface between general relativity, thermodynamics and quantum theory.

Since Hawking discovered that black holes emit a quasi-thermal radiation \cite{Hawking1975}, and therefore slowly evaporate away, a vast literature has studied the characteristics of this Hawking radiation. Following Hawking's seminal work, Teukolsky, Press, Page, Chandrasekhar, and Detweiler have worked out the equations governing the perturbations of rotating and charged Kerr--Newman BHs for perturbations with spins 0, 1, 2 and $1/2$ in general relativity. From this, they have then deduced the resulting rates of emission of Hawking radiation \cite{Teukolsky1,Teukolsky2,Teukolsky3,Page1,Page2,Page3,Chandra1,Chandra2,Chandra3,Chandra4} (see \cite{Chandrasekhar1983} for a complete mathematical review). Since then, it has been understood that general relativity is extremely likely to acquire corrections in both the infrared and ultraviolet regimes. These corrections naturally affect black hole physics. On the one hand, in the context of cosmology, general relativity has been challenged by the discovery of dark matter and dark energy and, for instance, the presence of a cosmological constant in the Einstein equations leads to anti-de Sitter types of BH metrics with modified Hawking radiation \cite{Hemming2001}. On the other hand, the attempts to reconcile general relativity with quantum theory have led to theories of quantum gravity extending general relativity into the deep quantum regime, such as string theory and loop quantum gravity. Although the purpose of such theories is to propose an ultraviolet completion of general relativity, there is a sense in which they naturally lead to observable effects at large scales. For instance string theory leads to extra spatial dimensions \cite{Harris2003,Johnson2020}, and loop quantum gravity leads to effective modifications in turn implying the avoidance of cosmological \cite{Ashtekar:2011ni} and black hole singularities \cite{Ashtekar:2005qt,Modesto:2005zm,Bohmer:2007wi,Ashtekar:2018lag,Bodendorfer:2019jay}. From this perspective, black holes act as probes and testbeds, translating the deep quantum corrections to the high curvature regime of general relativity within the horizon into semi-classical corrections to black hole properties as seen from outside the horizon.

Following this logic, various black hole solutions to the corrected Einstein equations in these many modified gravity frameworks have been proposed over the last decades, accompanied by the computation of the corresponding Hawking radiation. Recent work includes \textit{e.g.} massive gravity \cite{Kanzi2020}, cubic gravity \cite{Konoplya2020}, hairy BHs \cite{Chowdhury2020}, Einstein--Gauss--Bonnet BHs \cite{Zhang2020,Konoplya2020_2}, higher-dimensional BHs \cite{Harris2003,Johnson2020}, Kerr--Newman massive scalar emission \cite{Xu2020}, Hayward BHs \cite{Rincon2020,Molina2021}, general ``regular'' BHs \cite{Berry2021} and Kerr--Newman--de Sitter BHs \cite{Motohashi2021}. Beside the Hawking radiation, another important near-horizon property of black hole, which is very sensitive to modifications of general relativity and can be measured from the outside, is the detail of quasi-normal modes. They constitute the ringdown signal of a black hole relaxing towards its equilibrium state. This has become especially relevant in view of the recent gravitational wave detections from black hole mergers by LIGO/VIRGO (see \cite{LIGO2019} and references therein). Indeed, the increasing sensitivity of the gravitational wave detectors promises an access to the fine structure of the quasi-normal modes resulting from black hole mergers. Through this, we aim to push general relativity to its limits of validity. Indeed, there is (justified) hope that the measure of those  quasi-normal mode gravitational waves will give access to the precise characteristics of black hole horizons and thus to their correct metric description. Recent work has focused for example on charged Bardeen BHs \cite{Jusufi2020}, Gauss--Bonnet BHs \cite{Ma2018,Devi2020}, Palatini gravity \cite{Chen2018,Chen2019}, $f(R)$ gravity \cite{Soham2018}, Kerr--de Sitter BHs \cite{Novaes2019}, conformal gravity \cite{Chen2019_2,Chen:2021cts}, higher derivative gravity \cite{Cano2020}, and so-called polymerized BHs within loop quantum gravity \cite{Cruz:2015bcj,Barrau2019,Liu:2020ola}.

The computation of both Hawking radiation and quasi-normal modes is related to the response of black holes to perturbations. Thus, understanding the physically-measurable consequences of modified gravity on the Hawking radiation and quasi-normal modes requires to work out the equations of motion of the various spin perturbations to black hole metrics. This means generalizing the work of \cite{Teukolsky1,Teukolsky2,Teukolsky3,Page1,Page2,Page3,Chandra1,Chandra2,Chandra3,Chandra4,Chandrasekhar1983} to all black hole metrics predicted by the various modified gravity theories. In the present paper, we focus on spherically-symmetric static metrics of the form \eqref{eq:metric}, and show how the equations of motion can be written in a form similar to the Regge--Wheeler equation for Schwarzschild BHs, \emph{i.e.} as a one-dimensional Schr\"odinger-like radial wave equations with a short-ranged potential. This potential depends on the spin of the perturbation field and we give its explicit expression for each spin 0, 1, 2 and $1/2$. This derivation already exists in the literature for fields of spins 2 and 0  (see \emph{e.g.} respectively \cite{Barrau2019} and \cite{Hossenfelder2012}), but here we extend it to spins 1 and $1/2$. The potential for spin 1 already appears in \cite{Berry2021}, but however without the intermediate Teukolsky equation leading to the result (instead, the authors use an argument related to conformal invariance). Hawking radiation rates for spin $1/2$ fields in the case of polymerized BH metrics have been studied in \cite{Moulin2019}. As we will show in the forthcoming companion paper devoted to the detailed study of the Hawking spectra, our results differ from that of \cite{Moulin2019}. We believe that this is due to the fact that this reference does not use the analytic form of the spin $1/2$ short-ranged potential. This is the reason for which in the present paper we derive once and for all the analytic form of the Teukolsky equations and short-ranged potentials in the case of the general metrics \eqref{eq:metric}. We also give a general derivation of the intermediate Teukolsky equation for spins 0, 1, 2, $1/2$ and $3/2$ for these generalized metrics. This is given in formula \eqref{eq:teukolsky_general}. This result will be particularly useful in future work when studying the Hawking emission spectra for metrics of the general type \eqref{eq:metric}. In particular, the formulas for the Teukolsky equations and the short ranged potentials can be applied to the case of metrics with independent time and radial components ($F\neq G$ in the metric ansatz \eqref{eq:metric}), which is particularly interesting since these metrics arise as regular black holes in several effective models of quantum gravity \cite{Barrau2019,Moulin2019,Berry2021}.

We note that Kodama and Ishibashi (see \emph{e.g.}~\cite{Kodama2003}) develop a formalism that goes straight from the metric to the short range potentials, by means of the stress-energy tensor and without computing the intermediate Teukolsky equations. If it were to be extended to spins $1/2$ and $3/2$ as well as metrics with different time and radial components, this formalism would provide a complementary way of deriving the potentials. We have checked that our results are similar to those of~\cite{Cardoso2001}, to which \cite{Kodama2003} then compares (in the case of Schwarzschild-AdS BHs).

The paper is organized as follows. In section \ref{sec:starting_eqs}, we present the equations of motion using either a direct metric development or the Newman--Penrose formalism. Section \ref{sec:teukolsky} shows how to separate these equations to extract the one-dimensional radial Teukolsky equation for all spins. Section \ref{sec:potentials} presents the computation of the short-ranged potentials for all spins. In particular, the calculation for spins 1 and $1/2$ requires the use of a Chandrasekhar transform. Finally, section \ref{sec:examples} is devoted to the study of some examples of potentials for various black hole metrics, and their comparison with the Schwarzschild case. A mode detailed application of the formalism to various metrics will appear in the companion paper \cite{AAGLS2}.

\section{Metric and Newman--Penrose equations}
\label{sec:starting_eqs}

We consider spherically-symmetric static metrics, which constitute a subset of Petrov type D metrics. In four-dimensional Boyer--Lindquist coordinates, the general form of such metrics is
\begin{equation}\label{eq:metric}
	\d s^2=-G(r)\d t^2+\f{1}{F(r)}\d r^2+H(r)\d\Omega^2\,,
\end{equation}
where $\d\Omega^2=\d\theta^2+\sin\theta\,\d\varphi^2$ is the solid angle in spherical coordinates. Within this family of metrics, we further focus on solutions to the Einstein equations which are asymptotically flat. This means that at spatial infinity the functions $F$, $G$, and $H$ must satisfy the asymptotic conditions
\begin{equation}\label{falloffs}
	F(r)\underset{r\rightarrow+\infty}{\longrightarrow}1\,,\q G(r)\underset{r\rightarrow+\infty}{\longrightarrow}1\,,\q H(r)\underset{r\rightarrow+\infty}{\sim}r^2\,.
\end{equation}
Many usual metrics fall into this category, such as charged BHs, higher-dimensional BHs or effective BH metrics inspired by (loop) quantum gravity. One particular case that will be especially relevant is
\begin{equation}\label{eq:tr_symmetric}
	G(r)=F(r)\equiv h(r)\,,\q H(r)=r^2\,,
\end{equation}
to which we refer as $tr$-symmetric (for time-radius symmetric). For instance, charged and higher-dimensional BHs are $tr$-symmetric.

We now have to describe the dynamics of matter fields in these types of spacetimes. This can be done either by studying the equations of motion written in terms of the metric, or by using the Newman--Penrose formalism. In the following, we will use the most direct method to obtain the results.
Starting with the spin 0 case, we consider a massive scalar field $\phi$. In this case, it is easier to write the Proca equation in curved spacetime
\begin{equation}\label{eq:spin_0_NP}
	\big(\square+m_\phi^2\big)\phi=\f{1}{\sqrt{-g}}\partial_a\big(g^{ab}\sqrt{-g}\,\partial_b\phi\big)+m_\phi^2\phi=0\,,\q\sqrt{-g}=\sqrt{\f{G}{F}}H\sin\theta\,,
\end{equation}
where $m_\phi$ is the mass of the field.
For the other types of matter fields, the multiplicity of the vector, spinor or tensor components makes it difficult to obtain a single equation of motion when working directly with the metric. A simple and efficient way to bypass this difficulty is to exploit the Newman--Penrose formalism \cite{Newman1962,Chandrasekhar1983}, which relies on a reformulation of the equations of motion using a null tetrad field. A choice of null tetrad such that $g^{ab}=-l^an^b-n^al^b+m^a\bar{m}^b+\bar{m}^am^b$ is given by
\begin{equation}
	\begin{matrix*}[l]
	\displaystyle \phantom{n^a}l^a=\left(\f{1}{G},\sqrt{\f{F}{G}},0,0\right)\,, &
	\displaystyle m^a=\left(0,0,\f{1}{\sqrt{2H}},\f{i}{\sqrt{2H}\sin\theta}\right)\,,\vspace{0.2cm}\\
	\displaystyle \phantom{l^a}n^a=\left(\f{1}{2},-\f{\sqrt{FG}}{2},0,0\right)\,,\q
	&\displaystyle\bar{m}^a=\left(0,0,\f{1}{\sqrt{2H}},\f{-i}{\sqrt{2H}\sin\theta}\right)\,,
	\end{matrix*}\label{eq:tetrad}
\end{equation}
where $m$ and $\bar{m}$ are complex conjugate. This tetrad satisfies $l\cdot n=-1$ and $m\cdot\bar{m}=1$, while all other scalar products vanish. Introducing $e^a_i=(e^a_1,e^a_2,e^a_3,e^a_4)=(l^a,n^a,m^a,\bar{m}^a)$, we define the  $\lambda$-coefficients as
\begin{equation}
	\lambda_{ijk}\equiv\big(e_i^ae_k^b-e_k^ae_i^b\big)\partial_ae_{jb}\,.
\end{equation}
These coefficients enter the definition of the so-called Ricci spin (or rotation) coefficients
\begin{equation}
	\gamma_{ijk}\equiv\f{1}{2}(\lambda_{ijk}+\lambda_{kij}-\lambda_{jki})\,,
\end{equation}
and some specific linear combinations of these Ricci coefficients are then denoted by
\begin{equation}
\begin{matrix}
	\kappa\equiv\gamma_{311}\,,&\hspace{1cm}\rho\equiv\gamma_{314}\,,&\hspace{1cm}\epsilon\equiv(\gamma_{211}+\gamma_{341})/2\,,\\
	\sigma\equiv\gamma_{313}\,,&\hspace{1cm}\mu\equiv\gamma_{243}\,,&\hspace{1cm}\gamma\equiv(\gamma_{212}+\gamma_{342})/2\,,\\
	\lambda\equiv\gamma_{244}\,,&\hspace{1cm}\tau\equiv\gamma_{312}\,,&\hspace{1cm}\alpha\equiv(\gamma_{214}+\gamma_{344})/2\,,\\
	\nu\equiv\gamma_{242}\,,&\hspace{1cm}\pi\equiv\gamma_{241}\,,&\hspace{1cm}\beta\equiv(\gamma_{213}+\gamma_{343})/2\,.
\end{matrix}
\end{equation}
For the family of metrics \eqref{eq:metric}, the only non-vanishing components are real and given by
\begin{equation}
	\rho=-\f{H'}{2H}\sqrt{\f{F}{G}}\,,\q\mu=-\f{H'}{4H}\sqrt{FG}\,,\q\gamma=\f{G'}{4}\sqrt{\f{F}{G}}\,,\q\beta=-\alpha=\f{\cot\theta}{2\sqrt{2H}}\,,
\end{equation}
where $X'\equiv\partial_rX$ denotes the derivative in the radial direction. In the $tr$-symmetric case, these spin coefficients are the same as in \cite{Harris2003}.
We define the covariant derivatives along the four directions of the tetrad \eqref{eq:tetrad} as
\begin{equation}
	D\equiv l^a\nabla_a\,,\q\Delta\equiv n^a\nabla_a\,,\q\delta\equiv m^a\nabla_a\,,\q\bar{\delta}\equiv\bar{m}^a\nabla_a\,.
\end{equation}
These derivatives satisfy the general commutation relation
\begin{equation}
	\big(D-(p+1)\epsilon+q\rho+\bar{\epsilon}-\bar{\rho}\big)(\delta-p\beta+q\tau)=\big(\delta-(p+1)\beta+q\tau+\bar{\pi}-\bar{\alpha}\big)(D-p\epsilon+q\rho)\,,
\end{equation}
where $p$ and $q$ are arbitrary constants. This identity, which is valid for type D metrics (see equation (2.11) of \cite{Teukolsky1}),is pivotal in what follows. In particular, for the family of spherically-symmetric static metrics \eqref{eq:metric} that we focus on, it reduces to
\begin{equation}\label{eq:teukolsky_identity}
	(D+q\rho-\rho)(\delta+p\alpha)=(\delta+p\alpha)(D+q\rho)\,.
\end{equation}
We are now equipped with the necessary material to write down the Newman--Penrose equations of motion for fields of various spins.

\paragraph*{\textbf{Massless spin \boldsymbol{$1$}.}}

For a massless gauge boson, satisfying  the Einstein--Maxwell field equations $\d F=0$ and $\d*F=0$, the general form of the Newman--Penrose equations is \cite{Teukolsky1,Chandrasekhar1983,Harris2003}
\begin{subequations}
\begin{align}
	&D\phi_1-\bar{\delta}\phi_0+(2\alpha-\pi)\phi_0+\kappa\phi_2-2\rho\phi_1=0\,,\label{eq:NPME1}\\
	&D\phi_2-\bar{\delta}\phi_1+(2\epsilon-\rho)\phi_2+\lambda\phi_0-2\pi\phi_1=0\,,\\
	&\Delta\phi_0-\delta\phi_1-(2\gamma-\mu)\phi_0-\sigma\phi_2+2\tau\phi_1=0\,,\label{eq:NPME2}\\
	&\Delta\phi_1-\delta\phi_2-(2\beta-\tau)\phi_2-\nu\phi_0+2\mu\phi_1=0\,,
\end{align}
\end{subequations}
where the three Maxwell scalars are
\begin{equation}
	\phi_0\equiv F_{ab}l^am^b\,,\q\phi_1\equiv\f{1}{2}F_{ab}(l^an^b+\bar{m}^am^b)\,,\q\phi_2\equiv F_{ab}\bar{m}^an^b\,.
\end{equation}
The cancellation of many of the Ricci coefficients for the family of metrics \eqref{eq:metric} allows to write the first and third equations as a coupled system involving $\phi_0$ and $\phi_1$ only, \textit{i.e.}
\begin{subequations}
\begin{align}
	&(2\alpha-\bar{\delta})\phi_0+(D-2\rho)\phi_1=0\,,\\
	&(\Delta-2\gamma+\mu)\phi_0-\delta\phi_1=0\,.
\end{align}
\end{subequations}
These coupled first order equations can then be turn into a pair of decoupled second order differential equations. One applies $\delta$ to the first equation, and applies $D-3\rho$ to the second one. Adding the two resulting equations, and using the identity \eqref{eq:teukolsky_identity} with $p=0$ and $q=-2$, gives a differential equation involving $\phi_0$ only:
\begin{equation}\label{eq:spin_1_NP}
	\Big((D-3\rho)(\Delta-2\gamma+\mu)-\delta(\bar{\delta}-2\alpha)\Big)\phi_0=0\,.
\end{equation}
This is the equation of motion for a massless spin 1 field.

\paragraph*{\textbf{Massless spin \boldsymbol{$2$}.}}

For purely gravitational perturbations, which are equivalent to a massless spin 2 graviton field, the general form of the Newman--Penrose equations is \cite{Teukolsky1,Chandrasekhar1983}
\begin{subequations}
\begin{align}
	&(D-4\rho-2\epsilon)\psi_1-(\bar{\delta}-4\alpha+\pi)\psi_0+3\tilde{\kappa}\psi^\circ_2=0\,,\\
	&(\Delta-4\gamma+\mu)\psi_0-(\delta-4\tau-2\beta)\psi_1-3\tilde{\sigma}\psi^\circ_2=0\,,\\
	&(D-4\rho-\bar{\rho}-3\epsilon+\bar{\epsilon})\tilde{\sigma}\psi^\circ_2-(\delta-4\tau+\bar{\pi}-\bar{\alpha}-3\beta)\tilde{\kappa}\psi^\circ_2-\psi_0\psi^\circ_2=0\,,
\end{align}
\end{subequations}
where the $\psi_i$ are the perturbed components of the Weyl tensor (\textit{e.g.} $\psi_0\equiv-C_{abcd}l^am^bl^cm^d$), $\psi^\circ_2$ is the only non-vanishing background component, and the tilde on a spin coefficient indicates a perturbed quantity. We now specialize to the family of metrics \eqref{eq:metric}. If we remove the vanishing unperturbed spin coefficients, apply the operator $\delta-2\beta$ to the first equation, and the operator $D-5\rho$ to the second one, add the two and make use of identity \eqref{eq:teukolsky_identity} with $p=2$ and $q=-4$, we obtain an equation involving solely $\psi_0$, with the $\tilde{\sigma}\psi^\circ_2$ and $\tilde{\kappa}\psi^\circ_2$ contributions replaced by $\psi_0\psi^\circ_2$ thanks to the third equation. The resulting equation reads
\begin{equation}\label{eq:spin_2_NP}
	\Big((D-5\rho)(\Delta-4\gamma+\mu)-(\delta+2\alpha)(\bar{\delta}-4\alpha)-3\psi^\circ_2\Big)\psi_0=0\,,
\end{equation}
where the background $\psi^\circ_2$ is given by the Ricci identity as \cite{Chandrasekhar1983}
\begin{equation}\label{def Psi2}
	\psi^\circ_2=D\mu-\delta\pi-\bar{\rho}\mu-\sigma\lambda-\pi\bar{\pi}+(\epsilon+\bar{\epsilon})\mu+(\bar{\alpha}-\beta)\pi+\nu\kappa\qquad\Rightarrow\qquad\psi^\circ_2=D\mu-\rho\mu\,.
\end{equation}
Equation \eqref{eq:spin_2_NP} is the equation of motion for a massless spin 2 field.

\paragraph*{\textbf{Massless spin \boldsymbol{$1/2$}.}}

The Newman--Penrose equations for the massless Dirac spin $1/2$ field are \cite{Teukolsky1,Harris2003}
\begin{subequations}
\begin{align}
	&(\bar{\delta}-\alpha+\pi)\chi_0-(D-\rho+\epsilon)\chi_1=0\,,\\
	&(\Delta-\gamma+\mu)\chi_0-(\delta+\beta-\tau)\chi_1=0\,,
\end{align}
\end{subequations}
where $\chi_i$ are the two components of the spinor. We now specialize to the metrics \eqref{eq:metric}. We remove the vanishing spin coefficients, apply the operator $\delta-\alpha$ to the first equation, apply the operator $D-2\rho$ to the second one, subtract the two and make use of identity \eqref{eq:teukolsky_identity} with $p=-1$ and $q=-1$. This produces a decoupled differential equation for $\chi_0$ only:
\begin{equation}\label{eq:spin_12_NP}
	\Big((D-2\rho)(\Delta-\gamma+\mu)-(\delta-\alpha)(\bar{\delta}-\alpha)\Big)\chi_0=0\,.
\end{equation}
This is the equation of motion for a massless spin $1/2$ field.

\paragraph*{\textbf{Massless spin \boldsymbol{$3/2$}.}}

Finally, the general form of the Newman--Penrose equations for a Rarita--Schwinger massless spin $3/2$ field is \cite{Castillo1989}
\begin{subequations}
\begin{align}
	&(D-\epsilon-3\rho)H_{001}-(\bar{\delta}-3\alpha+\pi)H_{000}-\psi^\circ_2\psi_{000}=0\,,\\
	&(\delta-\beta-3\tau)H_{001}-(\Delta-3\gamma+\mu)H_{000}-\psi^\circ_2\psi_{001}=0\,,
\end{align}
\end{subequations}
where $H_{000}=(\delta-2\beta-\bar{\alpha}+\bar{\pi})\psi_{000}-(D-2\epsilon+\bar{\epsilon}-\bar{\rho})\psi_{001}$ is a combination of the spinor components, and $\psi^\circ_2$ is the same background component as in \eqref{def Psi2}. Specializing to the metric ansatz \eqref{eq:metric}, we remove the vanishing spin coefficients, apply the operator $\delta+\alpha$ to  the first equation, apply the operator $D-4\rho$ to the second one, subtract the two and use identity \eqref{eq:teukolsky_identity} with $p=1$ and $q=-3$. This leads to an equation on $H_0\equiv H_{000}$ only, which reads
\begin{equation}\label{eq:spin_32_NP}
	\Big((D-4\rho)(\Delta-3\gamma+\mu)-(\delta+\alpha)(\bar{\delta}-3\alpha)-\psi^\circ_2\Big)H_0=0\,,
\end{equation}
where we have also used $(D-3\rho)\psi^\circ_2=0$ and $(\delta-3\tau)\psi^\circ_2=0$, which follows from the Bianchi identities \cite{Teukolsky1}. Equation \eqref{eq:spin_32_NP} is the equation of motion for a massless spin $3/2$ field.

Following \cite{Li2011,Vagenas2020}, we note that equations \eqref{eq:spin_1_NP}, \eqref{eq:spin_2_NP}, \eqref{eq:spin_12_NP} and \eqref{eq:spin_32_NP} for $s>0$ fields can be recast under the remarkably compact form
\begin{align}
    &\big\{ \left[ D - (2s-1)\epsilon + \overline{\epsilon} - 2s\rho - \overline{\rho} \right]\left( \Delta - 2s\gamma + \mu \right) \nonumber\\
	&-\left[ \delta + \overline{\pi} - \overline{\alpha} - (2s-1)\beta - 2s\tau \right]\left( \overline{\delta} + \pi - 2s\alpha \right) \label{eq:master_NP} \\
	&-(2s-1)(s-1)\Psi_2\big\}\tilde{\Phi}_{s} = 0\,,\nonumber
\end{align}
with $\tilde{\Phi}_s = (\phi_0,\psi_0,\chi_0,H_0)$ depending on the spin.

We will now show how the various equations \eqref{eq:spin_0_NP}, \eqref{eq:spin_1_NP}, \eqref{eq:spin_2_NP}, \eqref{eq:spin_12_NP}, \eqref{eq:spin_32_NP} for all spins, or equivalently the equations \eqref{eq:spin_0_NP} and \eqref{eq:master_NP}, can be transformed into radial Teukolsky equations.

\section{Teukolsky equations for all spins}
\label{sec:teukolsky}

In this section we now derive an equivalent of the radial Teukolsky equation for all spins in the general spherically-symmetric and static metric \eqref{eq:metric}. The first step of this calculation consists in developing explicitly all the terms in equations \eqref{eq:spin_0_NP}, \eqref{eq:spin_1_NP}, \eqref{eq:spin_2_NP}, \eqref{eq:spin_12_NP}, \eqref{eq:spin_32_NP}. Then, based on the spherical and time symmetries of the metric \eqref{eq:metric}, we choose
\begin{equation}\label{eq:anzatz}
	\big(\phi,\phi_0,\psi_0,\chi_0,H_0\big)=\Phi_s(r)S^s_{\ell,m}(\theta,\varphi)e^{-i\omega t}\,,
\end{equation}
as an ansatz for the wavefunctions. Here $S^s_{\ell,m}$ are the spin-$s$ weighted spherical harmonics for angular modes $\ell,m$, satisfying the equation
\begin{equation}
	\left(\f{1}{\sin\theta}\partial_\theta(\sin\theta\,\partial_\theta)+\csc^2\theta\,\partial_\varphi^2+\f{2is\cot\theta}{\sin\theta}\partial_\varphi+s-s^2\cot^2\theta+\lambda_\ell^s\right)S_{\ell,m}^s=0\,,
\end{equation}
where the separation constant is $\lambda_\ell^s\equiv\ell(\ell+1)-s(s+1)$. In the spin 0 case, $S_{\ell,m}^0=Y_{\ell,m}$ are just the spherical harmonics. As we are here considering metrics with spherical and not axial symmetry, the dependency on the angular momentum projection $m$ factorizes as $S_{\ell,m}^s(\theta,\varphi)=S_\ell^s(\theta)e^{im\varphi}$. Expanding with \eqref{eq:anzatz} the equations of motion obtained above for all spins will now allow us to decouple the angular and radial equations, just like in the Schwarzschild and Kerr cases \cite{Teukolsky1,Teukolsky3}. Furthermore, the time symmetry replaces time derivatives by the energy $\omega$ of the field.

For the sake of clarity, we give the details of the calculations in appendix \ref{Teukolsky details}. The final result takes a remarkably simple form, and we obtain the one-dimensional radial Teukolsky equations \eqref{eq:teukolsky_0}, \eqref{eq:teukolsky_1}, \eqref{eq:teukolsky_2}, \eqref{eq:teukolsky_12}, and \eqref{eq:teukolsky_32}, for spins 0, 1, 2, $1/2$, and $3/2$ respectively. Equivalently, we can rewrite these results in the form of the master equation \eqref{eq:Teukolsky_master} valid for all spins. This radial Teukolsky equation \eqref{eq:Teukolsky_master} can be written in the general form
\begin{equation}\label{eq:teukolsky_general}
	A_s\big(B_s\Phi'_s\big)'+\left(\omega^2+i\omega s\sqrt{\f{F}{G}}\left(\f{GH'}{H}-G'\right)+C_s\right)\Phi_s=0\,,
\end{equation}
where the radial functions $A_s(r)$, $B_s(r)$, and $C_s(r)$ can be read in appendix \ref{Teukolsky details} for the various values of the spin $s$, and where once again a prime denotes the radial derivative. The consistency of this equation can be checked by choosing a $tr$-symmetric metric with \eqref{eq:tr_symmetric}. Inserting this in \eqref{eq:teukolsky_general} reproduces the Teukolsky master equation for all spins derived in \cite{Harris2003}, which is
\begin{equation}
	\f{1}{\Delta^s}\big(\Delta^{s+1}\Phi'_s\big)'+\left(\f{\omega^2r^2}{h}+2i\omega sr-\f{is\omega r^2h'}{h}+s(\Delta''-2)-\lambda_\ell^s\right)\Phi_s=0\,,
\end{equation}
where in \cite{Harris2003} the notation is $\Delta(r)\equiv r^2h(r)$.

\section{Short-ranged potentials}
\label{sec:potentials}

The next step towards an applicable formulation of the equations of motion, for the computation of both quasi normal modes and Hawking radiation, is to write the Teukolsky equations \eqref{eq:teukolsky_general} in the form of a Schr\"odinger wave equation with short-ranged potentials. Even if the equations can in principle be solved in the form \eqref{eq:teukolsky_general}, precise and stable numerical computations require to work with potentials which fall off at least as $1/r^2$ at infinity. Furthermore, working with real-valued potentials also constitutes an appreciable bonus. We therefore need to get rid of the first order radial derivatives and of the complex $is\omega$ terms in equation \eqref{eq:teukolsky_general}, which have a $1/r$ behaviour at infinity.

For all of this section, it will be convenient to define a generalized tortoise coordinate $r^*$ as \cite{Hossenfelder2012,Barrau2019}
\begin{equation}\label{eq:tortoise}
	\f{\d r^*}{\d r}=\f{1}{\sqrt{FG}}\,.
\end{equation}
In what follows we will give the expressions of the potentials with both the $r^*$ and $r$ coordinates, because the first one is more concise and the second one is better suited for numerical calculations. Furthermore, we will also consider the general redefinition of the wave function as
\begin{equation}\label{eq:transformation_1}
	\Psi_s\equiv\Phi_s\sqrt{\f{B_s}{\sqrt{FG}}}\,,
\end{equation}
where all quantities are functions of $r$ and we keep track of the spin $s$. Finally, for each spin our goal will be to find a wave function $Z_s$ satisfying the general Schr\"odinger-like equation
\begin{equation}
	\partial_*^2Z_s+\Big(\omega^2-V_s\big(r(r^*)\big)\Big)Z_s=0\,, \label{eq:Zeq}
\end{equation}
with spin-dependent potentials $V_s$, and where $\partial_*$ denotes the derivative with respect to the tortoise coordinate $r^*$. Spins 0 and 2 are already treated in the literature, while spins 1/2 and 1 require more work, and in particular the use of the Chandrasekhar transformation. We now study in detail these aspects.

\subsection{Spins 0 and 2}

For the massive spin 0 field, there is no complex term in \eqref{eq:teukolsky_0}, and all the terms are already decreasing faster than $1/r^2$ at infinity because of the fall-offs \eqref{falloffs}. Applying the transformations \eqref{eq:tortoise} and \eqref{eq:transformation_1}, we obtain simply a Schr\"odinger wave equation for $Z_0\equiv\Psi_0$, with a potential given by \cite{Hossenfelder2012}
\begin{align}
	V_0\big(r(r^*)\big)&=-Gm_\phi^2+\f{G\lambda_\ell^0}{H}+\f{1}{2}\sqrt{\f{FG}{H}}\left(\sqrt{\f{FG}{H}}H'\right)'\nonumber\\
	&=-Gm_\phi^2+\f{G\lambda_\ell^0}{H}+\f{\partial_*^2\sqrt{H}}{\sqrt{H}}\,.
\end{align}
This is the short-ranged potential for the massive spin 0 field in the metric \eqref{eq:metric}.

For the massless spin 2 field, reference \cite{Barrau2019} follows \cite{Chandrasekhar1983}. They consider clever combinations of the metric components and the vanishing of the Ricci tensor components at first order in the perturbation to obtain directly a decoupled radial equation of the Schr\"odinger-like form, with the potential
\begin{align}
	V_2\big(r(r^*)\big)&=\f{G(\lambda_\ell^2+4)}{H}+\f{FGH'^2}{2H^2}-\f{1}{2}\sqrt{\f{FG}{H}}\left(\sqrt{\f{FG}{H}}H'\right)'\nonumber\\
	&=\f{G(\lambda_\ell^2+4)}{H}+\f{(\partial_*H)^2}{2H^2}-\f{\partial_*^2\sqrt{H}}{\sqrt{H}}\,.
\end{align}
This is the short-ranged potential for the massless spin 2 field in the metric \eqref{eq:metric}.

\subsection{Spins 1 and 1/2}

We now complete the above results, which are already present in the literature, by deriving the short-ranged potentials for spins 1 and 1/2. These represent the main results of this article. In order to do so, we follow the method which has been used by Chandrasekhar and Detweiler to find the short-ranged potentials for the Kerr metric \cite{Chandra1,Chandra2,Chandra3,Chandra4}, and perform a Chandrasekhar transformation of the radial Teukolsky equations \eqref{eq:teukolsky_1} and \eqref{eq:teukolsky_12}.

Let us briefly look at the massless spin 1 and spin 1/2 fields separately before going back to general expressions for spin $s$. For the massless spin 1 field, applying the transformations \eqref{eq:tortoise} and \eqref{eq:transformation_1} to \eqref{eq:teukolsky_1} gives
\begin{align}\label{eq:spin_1_psi}
	&\left(\omega^2+i\omega\sqrt{\f{F}{G}}\left(\f{GH'}{H}-G'\right)+\f{FG''}{2}-\f{FGH''}{2H}-\f{FG'^2}{2G}+\f{FGH'^2}{4H^2}+\f{F'G'}{4}-\f{F'GH'}{4H}+\f{FG'H'}{4H}-\f{G(\lambda_\ell^1+2)}{H}\right)\Psi_1\nonumber\\
	&+\partial_*^2\Psi_1=0\,.
\end{align}
For the massless spin 1/2 field, \eqref{eq:teukolsky_12} becomes
\begin{align}\label{eq:spin_12_psi}
	&\left(\omega^2+i\omega\f{1}{2}\sqrt{\f{F}{G}}\left(\f{GH'}{H}-G'\right)+\f{FG''}{4}-\f{FGH''}{4H}-\f{3FG'^2}{16G}+\f{3FGH'^2}{16H^2}+\f{F'G'}{8}-\f{F'GH'}{8H}-\f{G(\lambda_{\ell}^{1/2}+1)}{H}\right)\Psi_{1/2}\nonumber\\
	&+\partial_*^2\Psi_{1/2}=0\,.
\end{align}
The general form of these equations is
\begin{equation}\label{eq:spin_s_psi}
	\left(\omega^2+i\omega s\sqrt{\f{F}{G}}\left(\f{GH'}{H}-G'\right)+D_s\right)\Psi_s+\partial_*^2\Psi_s=0\,.
\end{equation}
The only way to suppress the complex term without re-introducing first order derivatives is to change the unknown function $\Psi_s$ by a linear combination of itself and its first order derivative. In this context, this is called the Chandrasekhar transformation. In order to achieve this, we first define the intermediate function $Y_s$ by
\begin{equation}\label{eq:def_Y}
	\Psi_s=\alpha_sY_s\,.
\end{equation}
This function is such that equation \eqref{eq:spin_s_psi} can be written in the form
\begin{equation}\label{eq:def_Y_eq}
	\Lambda^2Y_s+P_s\Lambda_-Y_s-Q_sY_s=\partial_*^2Y_s+\omega^2Y_s+P_s(\partial_*Y_s+i\omega Y_s)-Q_sY_s=0\,,
\end{equation}
with two functions $P_s$ and $Q_s$, and the operators
\begin{equation}
	\Lambda_\pm\equiv\partial_*\pm i\sigma\,,\q\Lambda^2\equiv\Lambda_\pm\Lambda_\mp=\partial_*^2+\sigma^2\,,
\end{equation}
with $\sigma\equiv-\omega$. When written using \eqref{eq:def_Y}, equation \eqref{eq:spin_s_psi} becomes
\begin{equation}
	\partial_*^2Y_s+\omega^2Y_s+i\omega s\sqrt{\f{F}{G}}\left(\f{GH'}{H}-G'\right)Y_s+D_sY_s+\f{1}{\alpha_s}\big(2\partial_*\alpha_s\partial_*Y_s+Y_s\partial_*^2\alpha_s\big)=0\,.
\end{equation}
Comparing this result with \eqref{eq:def_Y_eq} then reveals that the two new functions are defined by the requirements
\begin{equation}\label{eq:def_P}
	Q_s=-D_s-\f{\partial_*^2\alpha_s}{\alpha_s}\,,\q P_s=\f{2\partial_*\alpha_s}{\alpha_s}=s\sqrt{\f{F}{G}}\left(\f{GH'}{H}-G'\right)=s\,\partial_*\ln\left(\f{H}{G}\right)\,.
\end{equation}
One can then show that this gives
\begin{equation}\label{eq:def_alpha_Q}
	Q_1=\f{G(\lambda_\ell^1+2)}{H}\,,\q Q_{1/2}=\f{G(\lambda_\ell^{1/2}+1)}{H}\,,\q\alpha_s=\left(\f{H}{G}\right)^{s/2}\,,
\end{equation}
where the expressions for $Q_1$ and $Q_{1/2}$ can explicitly be checked using \eqref{eq:spin_1_psi} and \eqref{eq:spin_12_psi}. Note that $Q_s$ takes a remarkably simple form, as displayed here, in the case of spin 1/2 and 1. Unfortunately, this is not true for spin 2 and 3/2, in which case the explicit expression is actually much more complicated. The solution for $\alpha_s$, however, is valid for all spins.

In order to continue with a lighter notation, from now on we remove the explicit spin label $s$ from all the various functions involved. We simply need to keep in mind that all the functions encountered below depend on the spin $s$. Now, let us further decompose $Y$ as a linear combination of the function $Z$ satisfying the Schr\"odinger wave equation \eqref{eq:Zeq}, by writing
\begin{equation}\label{eq:defY}
	Y\equiv f\Lambda_+\Lambda_+Z+W\Lambda_+Z\,,
\end{equation}
where on the right-hand side we have two unknown functions $f$ and $W$. The Schr\"odinger equation \eqref{eq:Zeq} takes the form $\Lambda^2Z=VZ$, where $V$ is the short-ranged potential that we are trying to determine for spin 1 and $1/2$. Acting on \eqref{eq:defY} with $\Lambda_-$ and using $\Lambda_+=\Lambda_-+2i\sigma$ then leads to
\begin{equation}\label{eq:defYminus}
	\Lambda_-Y=\big(\partial_*(f V)+WV\big)Z+\big(fV+\partial_*(W+2i\sigma f)\big)\Lambda_+Z\equiv-\f{\beta}{\alpha^2}Z+R\Lambda_+Z\,,
\end{equation}
where on the right-hand side we have introduced two unknown functions $\beta$ and $R$. Acting once again with $\Lambda_-$ on both sides gives
\begin{equation}
	\Lambda_-\Lambda_-Y=\left(2i\sigma\f{\beta}{\alpha^2}-\partial_*\left(\f{\beta}{\alpha^2}\right)+RV\right)Z+\left(\partial_*R-\f{\beta}{\alpha^2}\right)\Lambda_+Z\,.
\end{equation}
Next, we can use $\Lambda_+=\Lambda_-+2i\sigma$ once again to rewrite equation \eqref{eq:def_Y_eq} in the form
\begin{equation}
	\Lambda_-\Lambda_-Y=-(P+2i\sigma)\Lambda_-Y+QY=\left(\f{\beta}{\alpha^2}(P+2i\sigma)+QfV\right)Z+\big(Q(W+2i\sigma f)-(P+2i\sigma)R\big)\Lambda_+Z\,,
\end{equation}
where $P$ is given in equation \eqref{eq:def_P}. Matching the $Z$ and $\Lambda_+ Z$ terms of these two different expansions for $\Lambda_-\Lambda_-Y$ now tells us that we must have
\begin{align}\label{eq:Z_identification}
	RV-QfV=\f{\partial_*\beta}{\alpha^2}\,,\q
	\partial_*(\alpha^2R)=\beta+\alpha^2\big(Q(W+2i\sigma f)-2i\sigma R\big)\,,
\end{align}
in addition to which we should remember that, because of \eqref{eq:defYminus}, we also have the definitions
\begin{equation}\label{eq:useless}
	-\f{\beta}{\alpha^2}=\partial_*(fV)+WV\,,\q R=fV+\partial_*(W+2i\sigma f)\,.
\end{equation}
Now, one can check by a direct substitution that the four previous equations lead to the conservation equation
\begin{equation}\label{eq:integral}
	\partial_*\big(\alpha^2RfV+\beta(W+2i\sigma f)\big)=0\,,
\end{equation}
which is a generalization of Chandrasekhar's result \cite{Chandra1,Chandra2,Chandra3,Chandra4}. We call this constant $K$, and we will see later on that it simplifies the calculations neatly. We also define the quantity $T\equiv W+2i\sigma$. Using the identity \eqref{eq:integral} to remove an unwanted derivative of the potential $V$ (which would have caused further difficulties), we finally obtain that \eqref{eq:Z_identification} and \eqref{eq:useless} reduce to the following system of four equations:
\begin{subequations}\label{eq:system}
\begin{align}
	&RV-QfV=\f{\partial_*\beta}{\alpha^2}\,,\label{eq:fine_4}\\
	&\partial_*(\alpha^2R)=\beta+\alpha^2(QT-2i\sigma R)\,,\label{eq:fine_2}\\
	&R(R-\partial_*T)+\f{\beta T}{\alpha^2}=\f{K}{\alpha^2}\,,\label{eq:fine_3}\\
	&R=fV+\partial_*T\,,\label{eq:fine_1}
\end{align}
\end{subequations}
where \eqref{eq:fine_3} has been obtained by combining \eqref{eq:useless} and \eqref{eq:integral}. This is the system that we have to solve in order to prove that a solution $Z$ satisfying the Schr\"odinger wave equation in the potential $V$ does indeed exist. This system follows from the form of the Chandrasekhar transformation, and is valid for all spins.\footnote{We remember that in all the functions appearing in this system we have kept the spin label $s$ implicit for conciseness.} Chandrasekhar and Detweiler have solved it for the Kerr metric and for spins 0, 1, 2 and $1/2$. We will now solve it in the general case of the metric \eqref{eq:metric} for spin 1 following \cite{Chandra1}, and for spin $1/2$ following \cite{Chandra4}.

\subsubsection*{Spin $1$}

In the case of spin 1 we look for a simple solution, \emph{i.e.} we suppose that the unknown quantities are linear in $\sigma$ and of the form $A=A_1+2i\sigma A_2$, and that the desired potential $V$ is of course independent of $\sigma$ (together with $Q$ which is the initial potential without the $i\omega$ part). Looking at the system \eqref{eq:system} tells us that the only $\sigma^2$ term will come from $R_2$, meaning that we need to actually choose $R_2=0$. Then, if we do not wish to carry out the integrations, we further assume that $\partial_*T_2=0$. The only remaining term in $i\sigma$ then come from $f_2$ and $\partial_*\beta_2$, so we take $f_2=0$ and $\beta_2$ to be constant. Indeed as in \cite{Chandra1}, both $R$ and $f$ are also independent of $\sigma$ with these hypotheses. We therefore only need to decompose
\begin{equation}
	T\equiv T_1+2i\sigma T_2\,,\q K\equiv K_1+2i\sigma K_2\,,\q\beta\equiv\beta_1+2i\sigma\beta_2\,.
\end{equation}
With all these assumptions, the system \eqref{eq:system} simply becomes
\begin{subequations}
\begin{align}
	&RV-QfV=\f{\partial_*\beta_1}{\alpha^2}\label{eq:further_4}\,,\\
	&\partial_*(\alpha^2R)=\beta_1+2i\sigma\beta_2+\alpha^2\big(Q(T_1+2i\sigma T_2)-2i\sigma R\big)\,,\label{eq:further_2} \\
	&R(R-\partial_*T_1)+\f{1}{\alpha^2}\big(\beta_1T_1+2i\sigma(\beta_1T_2+\beta_2T_1)-4\sigma^2\beta_2T_2\big)=\f{1}{\alpha^2}(K_1+2i\sigma K_2)\,,\label{eq:further_3}\\
	&R=fV+\partial_*T_1\,.\label{eq:further_1}
\end{align}
\end{subequations}
Identifying the no-$\sigma$ and $\sigma$ terms in \eqref{eq:further_2} then gives us the two equations
\begin{equation}\label{eq:dR_RFdef}
	\partial_*(\alpha^2R)=\beta_1+\alpha^2QT_1\,,\q R=\f{\beta_2}{\alpha^2}+QT_2\,,
\end{equation}
while doing the same in \eqref{eq:further_3} leads to
\begin{equation}\label{eq:defbeta1}
	R(R-\partial_*T_1)+\f{1}{\alpha^2}\big(\beta_1T_1-4\sigma^2\beta_2T_2\big)=\f{K_1}{\alpha^2}\,,\q\beta_2T_1+\beta_1T_2=K_2\,.
\end{equation}
We can see from the last equation that the numerical value of the constant $T_2$ can be absorbed in the other unknown quantities, so we set $T_2=1$ and define $\kappa\equiv K_1+4\sigma^2\beta_2$. In order to rewrite the system in an elegant way, we now define the function
\begin{equation}
\mathcal{F}\equiv\alpha^2Q=\ell(\ell+1)\,.
\end{equation}
With this, the second equation in \eqref{eq:dR_RFdef} gives
\begin{equation}\label{eq:Ruseless}
	\alpha^2R=\beta_2+\mathcal{F}\,,
\end{equation}
which can then be injected in the first equation of \eqref{eq:dR_RFdef} to find
\begin{equation}
	\partial_*\mathcal{F}=\beta_1+T_1\mathcal{F}\,.
\end{equation}
We can now use \eqref{eq:defbeta1} to eliminate $\beta_1$ from all other equations, and \eqref{eq:Ruseless} to eliminate $R$. We can then write the previous equation as
\begin{equation}\label{eq:T1def}
	T_1=\f{1}{\mathcal{F}-\beta_2}(\partial_*\mathcal{F}-K_2)\,,
\end{equation}
and \eqref{eq:further_3} can be rewritten in the form
\begin{equation}
	\f{1}{\alpha^2}(\mathcal{F}+\beta_2)^2-(\mathcal{F}+\beta_2)\partial_*T_1+T_1(K_2-\beta_2T_1)=\kappa\,.
\end{equation}
Substituting the expression \eqref{eq:T1def} for $T_1$, we finally obtain an identity on $\mathcal{F}$ which reads \cite{Chandra1}
\begin{equation}\label{eq:identityF}
	\mathcal{F}(\partial_*\mathcal{F})^2+(\beta_2^2-\mathcal{F}^2)\partial_*^2\mathcal{F}+\f{\mathcal{F}^4}{\alpha^2}-\left(\f{2\beta_2^2}{\alpha^2}+\kappa \right)\mathcal{F}^2+(2\kappa\beta_2-K_2^2)\mathcal{F}+\left(\f{\beta_2^4}{\alpha^2}-\kappa\beta_2^2\right)=0\,.
\end{equation}
The goal is now to find a set of constants $\beta_2$, $\kappa$, and $K_2$ compatible with this identity. Since $\mathcal{F}$ is a constant given simply by $\mathcal{F}=\ell(\ell+1)$, this is actually straightforward. We deduce that $\beta_2=\pm\ell(\ell+1)$ and $K_2=0$, while $\kappa$ is unconstrained. We then choose $K_1=0$ and obtain $\kappa=4\sigma^2\beta_2$. Finally, since \eqref{eq:T1def} does not constrain $T_1$, we choose $T_1=0$, which implies $\beta_1=0$ thanks to \eqref{eq:defbeta1}. We have therefore found a consistent set of constants satisfying the assumptions, and all the remaining functions can be analytically computed. At the end, equations \eqref{eq:further_4} and \eqref{eq:further_1} lead to the very simple result
\begin{equation}
	V_1\big(r(r^*)\big)=Q_1=\ell(\ell+1)\f{G}{H}\,.
\end{equation}
This is the short-ranged potential for a massless spin 1 field in the metric \eqref{eq:metric}, which is moreover coherent with the potential obtained in \cite{Berry2021}.

\subsubsection*{Spin $1/2$}

In order to study the case of spin 1/2, we first note that the definition of $\mathcal{F}$ gives the simple result
\begin{equation}
	\mathcal{F}=\alpha^2Q=(\lambda_{\ell}^{1/2}+1)\sqrt{\f{F}{G}}\,.
\end{equation}
In spite of this simple form, using the same hypothesis as in the previous subsection, leading to \eqref{eq:identityF}, we find that the latter has no solution. We therefore need to make fewer assumptions than above. We will in fact follow \cite{Chandra4}, and go back to the system of equations \eqref{eq:system}. In this system, integrations can be avoided by assuming
\begin{equation}
	\partial_*T=0\,,\q\partial_*(\alpha^2R)=0\,,
\end{equation}
which in turn implies that $\tilde{R}\equiv\alpha^2R$ is a constant. Thus we have the system
\begin{subequations}\label{eq:system12}
\begin{align}
	&V=\f{\partial_*\beta}{\tilde{R}}+\f{(\lambda_{\ell}^{1/2}+1)}{\alpha^4}\,,\label{eq:spin12_fine_4}\\
	&0=\beta+\f{(\lambda_{\ell}^{1/2}+1)T}{\alpha^2}-2i\sigma \tilde{R}\,,\label{eq:spin12_fine_2}\\
	&\f{\tilde{R}^2}{\alpha^4}+\f{\beta T}{\alpha^2}=\f{K}{\alpha^2}\,,\label{eq:spin12_fine_3}\\
	&\f{\tilde{R}}{\alpha^2}=fV\,,\label{eq:spin12_fine_1}
\end{align}
\end{subequations}
where, in order to obtain \eqref{eq:spin12_fine_4}, we have used \eqref{eq:spin12_fine_1}. We see that \eqref{eq:spin12_fine_4} already gives us the potential as a function of $\beta$ and $\tilde{R}$. The goal is therefore to determine these functions. For this, we set $T=2i\sigma$ by analogy with the final result of \cite{Chandra4} and the result of the spin 1 calculation. Equation \eqref{eq:spin12_fine_2} then becomes
\begin{equation}
	\beta=2i\sigma\left(\tilde{R}-\f{(\lambda_{\ell}^{1/2}+1)}{\alpha^2}\right)\,,
\end{equation}
and \eqref{eq:spin12_fine_3} gives
\begin{equation}
	\tilde{R}^2+4\sigma^2(\lambda_{\ell}^{1/2}+1)=\alpha^2(4\sigma^2\tilde{R}+K)\,.
\end{equation}
In this equation, everything is a constant except $\alpha$, we can therefore identify separately
\begin{equation}
	K=-4\sigma^2\tilde{R}\,,\q\tilde{R}=\pm 2i\sigma\sqrt{(\lambda_{\ell}^{1/2}+1)}\,.
\end{equation}
We have therefore found a set of constants and a function $\beta$ satisfying the system \eqref{eq:system12}. We can finally use \eqref{eq:spin12_fine_4} to write the potential as
\begin{align}
	V_{1/2}\big(r(r^*)\big)&=\big(\ell(\ell+1)+1/4\big)\f{G}{H} \pm \sqrt{\ell(\ell+1)+1/4}\,\partial_*\left(\sqrt{\f{G}{H}}\right)\nonumber\\
	&=\big(\ell(\ell+1)+1/4\big)\f{G}{H}\pm\sqrt{\ell(\ell+1)+1/4}\,\sqrt{FG}\left(\sqrt{\f{G}{H}}\right)'\,.
\end{align}
This is the short-ranged potential for the massless spin 1/2 field in the metric \eqref{eq:metric}.

\subsection{Summary}

In this section we have obtained the short-ranged massless potentials for all spins in elegant forms using the tortoise coordinate $r^*$. This is summarized as
\begin{subequations}\label{eq:potentials}
\begin{align}
	&V_0=\nu_0\f{G}{H}+\f{\partial_*^2\sqrt{H}}{\sqrt{H}}\,,\\
	&V_1=\nu_1\f{G}{H}\,,\\
	&V_2=\nu_2\f{G}{H}+\f{(\partial_*H)^2}{2H^2}-\f{\partial_*^2\sqrt{H}}{\sqrt{H}}\,,\\
	&V_{1/2}=\nu_{1/2}\f{G}{H} \pm \sqrt{\nu_{1/2}}\,\partial_*\left( \sqrt{\f{G}{H}} \right)\,,
\end{align}
\end{subequations}
where we have defined for conciseness $\nu_0\equiv\ell(\ell+1)\equiv\nu_1$, $\nu_2\equiv\ell(\ell+1)-2$ and $\nu_{1/2}\equiv\ell(\ell+1)+1/4$. These results are surprisingly compact, and extend the existing literature to the case of spin 1 and 1/2 massless fields in the metric \eqref{eq:metric}. We can now focus on specific examples of metrics, and see what the resulting potentials look like when compared to the Schwarzschild ones.

\section{Some examples}
\label{sec:examples}

There are numerous physically-motivated spherically-symmetric and static metrics of the form \eqref{eq:metric}. These examples come from both classical general relativity and modified theories of gravity with \textit{e.g.} quantum gravity corrections. In this section we will study three examples of potentials, for charged BHs, higher-dimensional BHs, and finally so-called polymerized BHs with corrections from loop quantum gravity. We will have a more extensive discussion of the applications of the formalism presented here in the companion paper \cite{AAGLS2}.

There is already important physical information which can be extracted directly from the form of the potentials given below. For example, BHs which have a higher potential barrier than in the Schwarzschild case will have a lower Hawking emission rate. While this cannot directly be seen from the plots below because they are rescaled, it can easily be seen by looking at the analytic form of the potentials.

\subsection{${tr}$-symmetric case}

Charged and higher-dimensional BHs fall within the family of $tr$-symmetric metrics \eqref{eq:tr_symmetric}. We therefore first give general results about this case. Using a $tr$-symmetric ansatz, which depends only on a single function $h(r)$, the massless potentials \eqref{eq:potentials} become
\begin{subequations}
\begin{align}
	&V_0=h\left(\f{\ell(\ell+1)}{r^2}+\f{1}{r}h'\right)\,,\\
	&V_1=h\f{\ell(\ell+1)}{r^2}\,,\\
	&V_2=h\left(\f{\ell(\ell+1)}{r^2}-\f{1}{r}h'+\f{2(h-1)}{r^2}\right)\,,\\
	&V_{1/2}=h\f{\ell(\ell+1)+1/4}{r^2}\pm h^{1/2}\f{\sqrt{\ell(\ell+1)+1/4}}{r}h'\mp h^{3/2}\f{\sqrt{\ell(\ell+1)+1/4}}{r^2}\,.
\end{align}
\end{subequations}
The first three potentials, which are bosonic, can be written as a single master potential
\begin{equation}
	V_s=h\left(\f{\ell(\ell+1)}{r^2}+\f{1-s}{r}h'+\f{s(s-1)(h-1)}{r^2}\right)\,.
\end{equation}
We note that the last term in this master potential is absent from equation (6) of \cite{Rincon2020}, but is coherent with our results and with that of \cite{Barrau2019}.\footnote{We therefore conclude that the master equation of \cite{Rincon2020} is not valid for spin 2.} For the Schwarzschild metric, we recall that
\begin{equation}\label{eq:Schwarschild}
	F=G=h=1-\f{r\mrm{S}}{r}\,,\q H=r^2\,,
\end{equation}
where $r\mrm{S} = 2M \equiv r\mrm{H}$ is the Schwarzschild radius of the horizon.

\subsection{Charged black holes}

\begin{figure}[ht]
	\centering
	\includegraphics[scale=0.75]{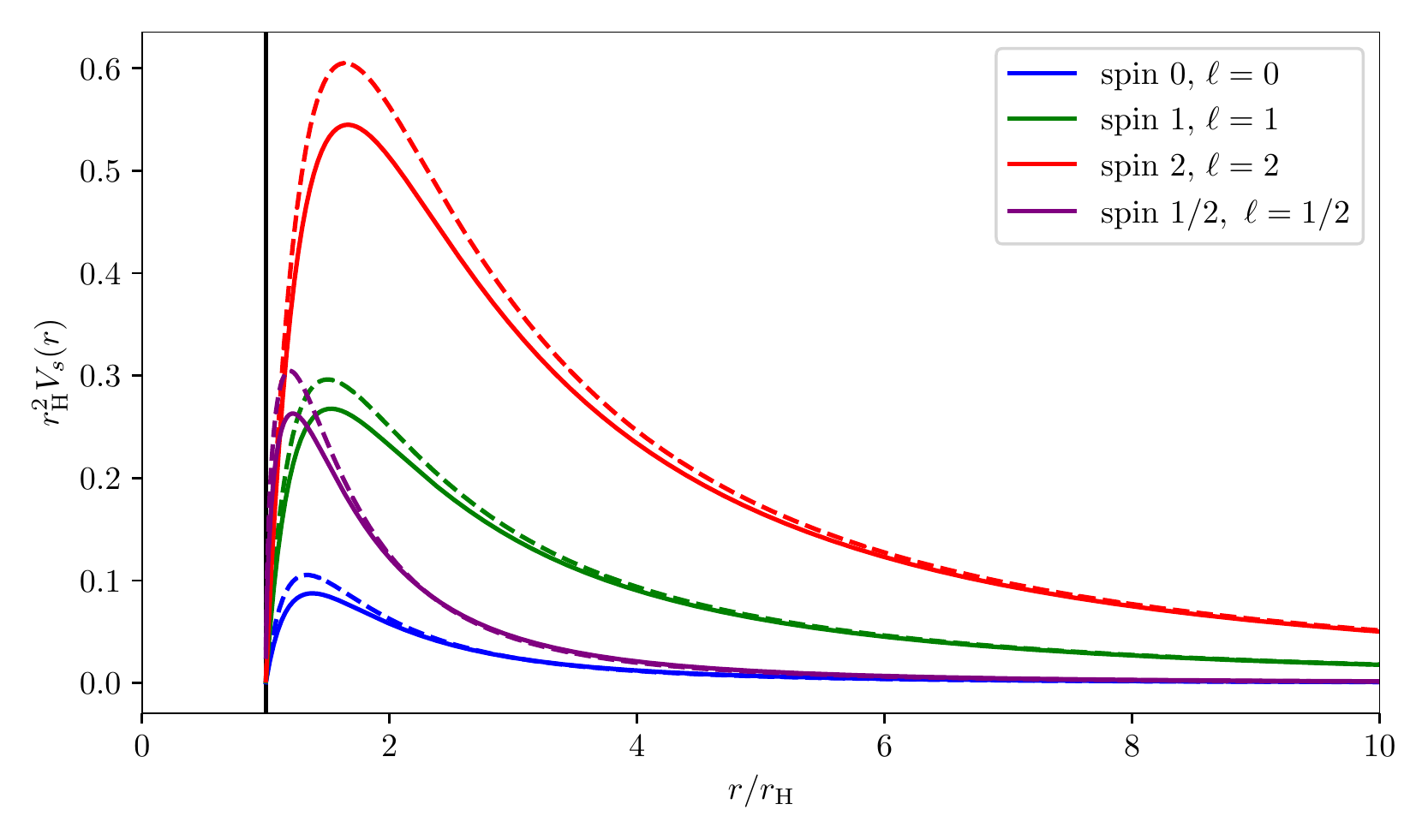}
	\caption{Comparison of the potentials for a charged BH with $r\mrm{Q}=r\mrm{S}/3$ (solid lines) and for the Schwarzschild metric (dashed lines). The vertical black line represents the BH horizon.}
	\label{fig:charged}
\end{figure}

After the Schwarzschild solution \eqref{eq:Schwarschild}, the simplest $tr$-symmetric physically-relevant BH which is solution of classical general relativity equations is the charged BH with
\begin{equation}
	F=G=h=1-\f{r\mrm{S}}{r}+\f{r\mrm{Q}^2}{r^2}\,,\q H=r^2\,,
\end{equation}
where $r\mrm{Q}^2\equiv Q^2$ and $Q<M$ is the charge of the BH (since we are working with natural units $4\pi\varepsilon_0=1$, the fine structure constant is $\alpha_{\rm em}=1$). The exterior horizon is given by
\begin{equation}
    r\mrm{H} \equiv r_+ = r\mrm{S}\f{1+\sqrt{1-4r\mrm{Q}^2/r\mrm{S}^2}}{2}\,.
\end{equation}
For neutral particles (\textit{i.e.} with no additional coupling between the charge of the BH and that of the particle), the potentials take the form
\begin{subequations}
\begin{align}
	&V_0=\f{\nu_0}{r^2}+\f{(1-\nu_0)r\mrm{S}}{r^3}+\f{r\mrm{Q}^2(\nu_0-2)-r\mrm{S}^2}{r^4}+\f{r\mrm{Q}^2r\mrm{S}}{r^5}-\f{2r\mrm{Q}^4}{r^6}\,,\\
	&V_1=\f{\nu_1}{r^2}-\f{\nu_1r\mrm{S}}{r^3}+\f{\nu_1r\mrm{Q}^2}{r^4}\,,\\
	&V_2=\f{\nu_2+2}{r^2}-\f{(\nu_2+3)r\mrm{S}}{r^3}+\f{(\nu_2+4)r\mrm{Q}^2+r\mrm{S}^2}{r^4}-\f{r\mrm{Q}^2 r\mrm{S}}{r^5}+\f{2r\mrm{Q}^4}{r^6}\,,\\
	&V_{1/2}=\f{\nu_{1/2}}{r^2}-\f{\nu_{1/2}r\mrm{S}}{r^3}+\f{\nu_{1/2}r\mrm{Q}^2}{r^4}\mp\f{\sqrt{\nu_{1/2}}}{2}r\sqrt{1-\f{r\mrm{S}}{r}+\f{r\mrm{Q}^2}{r^2}}\left(\f{2}{r^3}-\f{3r\mrm{S}}{r^4}+\f{4r\mrm{Q}^2}{r^5}\right)\,.
\end{align}
\end{subequations}
In figure \ref{fig:charged} we show these potentials compared to the Schwarzschild ones for the minimum possible angular momenta $\ell=s$ and for $r\mrm{Q}=r\mrm{S}/3$ (that is to say $Q=2M/3$).

\subsection{Higher-dimensional black holes}

\begin{figure}[ht]
    \centering
    \includegraphics[scale=0.75]{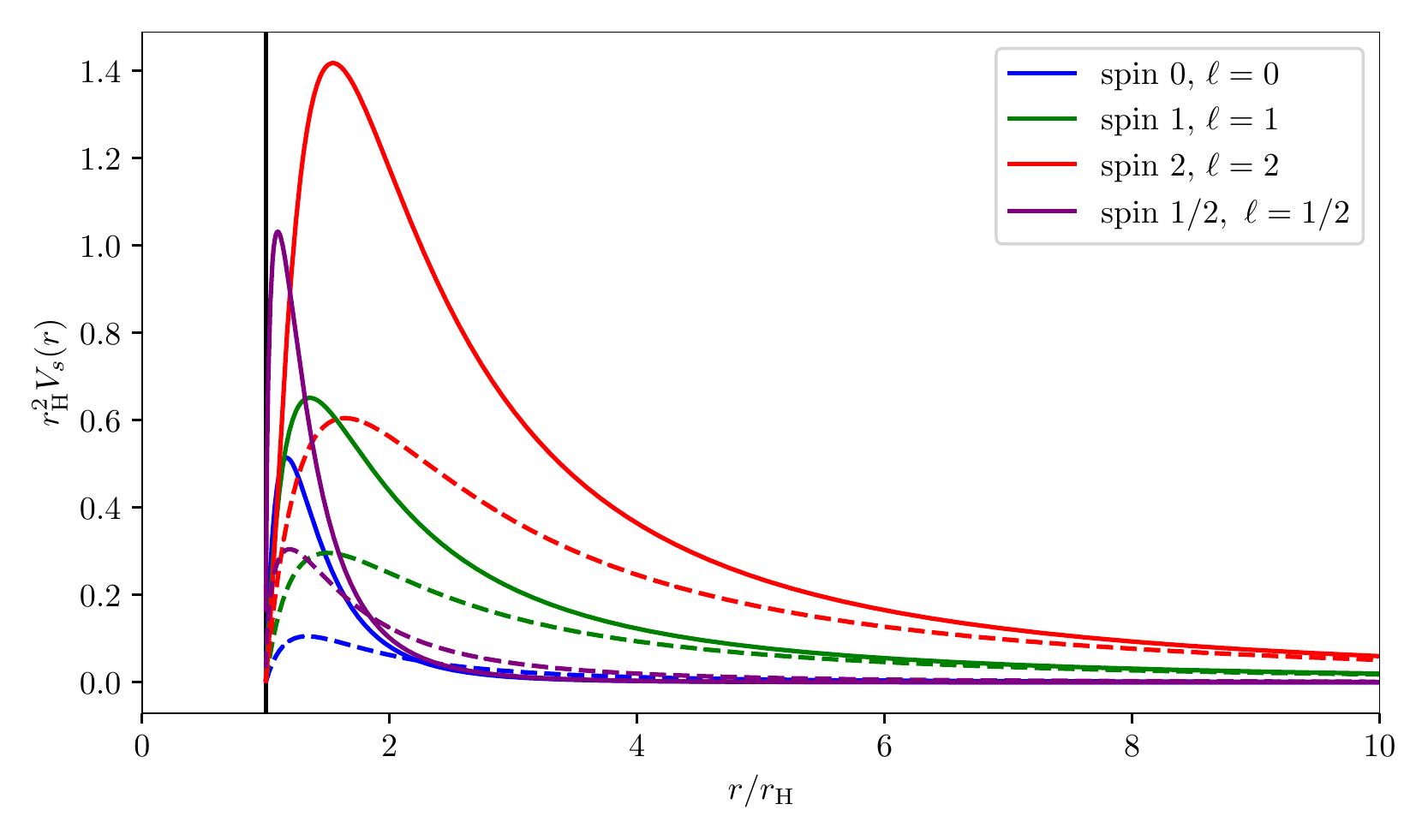}
    \caption{Comparison of the potentials for a higher-dimensional BH with $n=2$, $M=10^{10}\,M\mrm{P}$ and $M_*=10\,$TeV (solid lines) and for the Schwarzschild metric (dashed lines). The vertical black line denotes the BH horizon.}
    \label{fig:higher}
\end{figure}

Another simple case of $tr$-symmetric metrics describes $(4+n)$-dimensional BHs \cite{Harris2003,Johnson2020}. In this case the geometry is specified by
\begin{equation}
	F=G=h\equiv1-\left(\f{r\mrm{H}}{r}\right)^{n+1}\,,\q H=r^2\,,
\end{equation}
where the horizon radius is\footnote{\cite{Harris2003} assumes that these BHs satisfy $\ell\mrm{P}\ll r\mrm{H}\ll R$ where $\ell\mrm{P}$ is the Planck length and $R$ is the typical size of the extra dimensions.}
\begin{equation}
	r\mrm{H}=\f{1}{\sqrt{\pi}M_*}\left(\f{M}{M_*}\right)^{1/(n+1)}\left(\f{8\Gamma\big((n+3)/2\big)}{n+2}\right)^{1/(n+1)}\,,
\end{equation}
and $M\mrm{P}^2\sim M_*^{n+2}R^n$ defines the fundamental mass scale of the theory. In this geometry, the massless potentials become
\begin{subequations}
\begin{align}
	&V_0=\f{\nu_0}{r^2}+\f{r\mrm{H}^{n+1}(n+1-\nu_0)}{r^{n+3}}-\f{(n+1)r\mrm{H}^{2n+2}}{r^{2n+4}}\,,\\
	&V_1=\f{\nu_1}{r^2}-\f{\nu_1 r\mrm{H}^{n+1}}{r^{n+3}}\,,\\
	&V_2=\f{\nu_2+2}{r^2}-\f{\big(\nu_2+2+(n+1)\big)r\mrm{H}^{n+1}}{r^{n+3}}+\f{(n+1)r\mrm{H}^{2n+2}}{r^{2n+4}}\,,\\
	&V_{1/2} = \f{\nu_{1/2}}{r^2}-\f{\nu_{1/2}r\mrm{H}^{n+1}}{r^{n+3}} \mp \f{\sqrt{\nu_{1/2}}}{2}\sqrt{1-\left( \f{r\mrm{H}}{r} \right)^{n+1}}\left(\f{2}{r^2}-\f{(n+3)r\mrm{H}^{n+1}}{r^{n+3}} \right)\,.
\end{align}
\end{subequations}
Note that these potentials describe the radiation truncated to the four-dimensional $(t,r,\theta,\varphi)$ subspace, and in particular does not describe the radiation within the extra dimensions. In figure \ref{fig:higher} we plot these potentials and compare them to the Schwarzschild ones for $n=2$, $M=10^{10}\,M\mrm{P}$ and $M_*=10\,$TeV \cite{Johnson2020}.

\subsection{Polymerized black holes}

Interesting metrics which are not $tr$-symmetric arise in loop quantum gravity, where effective semi-classical corrections due to effects of quantum gravity have been derived and give rise to so-called polymerized BHs. There are many proposals for deriving such BH metrics \cite{Modesto:2005zm,Bohmer:2007wi,Ashtekar:2018cay,BenAchour:2018khr,Bojowald:2018xxu,Bodendorfer:2019cyv,Bodendorfer:2019jay}, as we will review in the companion paper \cite{AAGLS2}. Here, for the sake of the example and in order to compare with previous results obtained in \cite{Hossenfelder2012,Barrau2019}, we will focus on the particular type of polymerized BHs with \cite{Modesto:2008im}
\begin{equation}
	F=\f{(r-r_+)(r-r_-)r^4}{(r+r_*)^2(r^4+a_0^2)}\,,\q G=\f{(r-r_+)(r-r_-)(r+r_*)^2}{r^4+a_0^2}\,,\q H=r^2+\f{a_0^2}{r^2}\,.
\end{equation}
Here $a_0$ is the area gap of loop quantum gravity, and the radii are given by
\begin{equation}
	r_+=2m\equiv r\mrm{H}\,,\q r_-=2mP^2\,,\q r_*=\sqrt{r_+r_-}\,,
\end{equation}
where $P=(\sqrt{1+\epsilon^2}-1)/(\sqrt{1+\epsilon^2}+1)$ is the so-called polymeric function, and the parameter $m$ is related to the so-called ADM mass $M$ by $M=m(1+P)^2$. These polymerized BH solutions therefore have two free parameters, which are $a_0$ and $\epsilon$. With these ingredients, the massless potentials become
\begin{subequations}
\begin{align}
	&V_0=\f{(r-r_+)(r-r_-)}{(r^4+a_0^2)^4}\Big(\nu_0r^{12}+(2\nu_0 r_*+r_++r_-)r^{11}+(\nu_0-2)r_*^2 r^{10}+2a_0^2(\nu_0+5)r^8\nonumber \\
	&\phantom{V_0=}+2a_0^2\big(2\nu_0r_*-5(r_++r_-)\big)r^7+2a_0^2r_*^2(\nu_0+5)r^6+ a_0^4(\nu_0-2)r^4+a_0^4(2\nu_0 r_*+r_++r_-)r^3+a_0^4 \nu_0 r_*^2 r^2\Big)\,,\\
	&V_1=\nu_1\f{r^2(r-r_+)(r-r_-)(r+r_*)^2}{(r^4+a_0^2)^2}\,,\\
	&V_2=\f{(r-r_+)(r-r_-)}{(r^4+a_0^2)^4}\Big((\nu_2+1)r^{12}+(2\nu_2r_*+r_++r_-)r^{11}+(\nu_2+2)r_*^2 r^{10}+a_0^2(2\nu_2-11)r^8\nonumber \\
	&\phantom{V_2=}+ 2a_0^2\big(\nu_2 r_*+5(r_++r_-)\big)r^7+a_0^2r_*^2(\nu_2-10)r^6+a_0^4(\nu_2+1)r^4+a_0^4(2\nu_2r_*-r_+-r_-)r^3+\nu_2 a_0^4 r_*^2 r^2+a_0^6\Big)\,,\\
	&V_{1/2}=\nu_{1/2}\f{r^2(r-r_+)(r-r_-)(r+r_*)^2}{(r^4+a_0^2)^2}\pm\f{\sqrt{\nu_{1/2}}}{2}\f{r\sqrt{(r -r_+)(r-r_-)}}{(r^4+a_0^2)^3}\Big((r^4+a_0^2)\Big[r^2(r+r_*)(2r-r_+-r_-)\nonumber \\
	&\phantom{V_{1/2}=}+2r^2(r-r_+)(r-r_-)+2r(r-r_+)(r-r_-)(r+r_*)\Big]-8r^5(r-r_+)(r-r_-)(r+r_*)\Big)\,.
\end{align}
\end{subequations}

\begin{figure}[ht]
    \centering
    \includegraphics[scale=0.75]{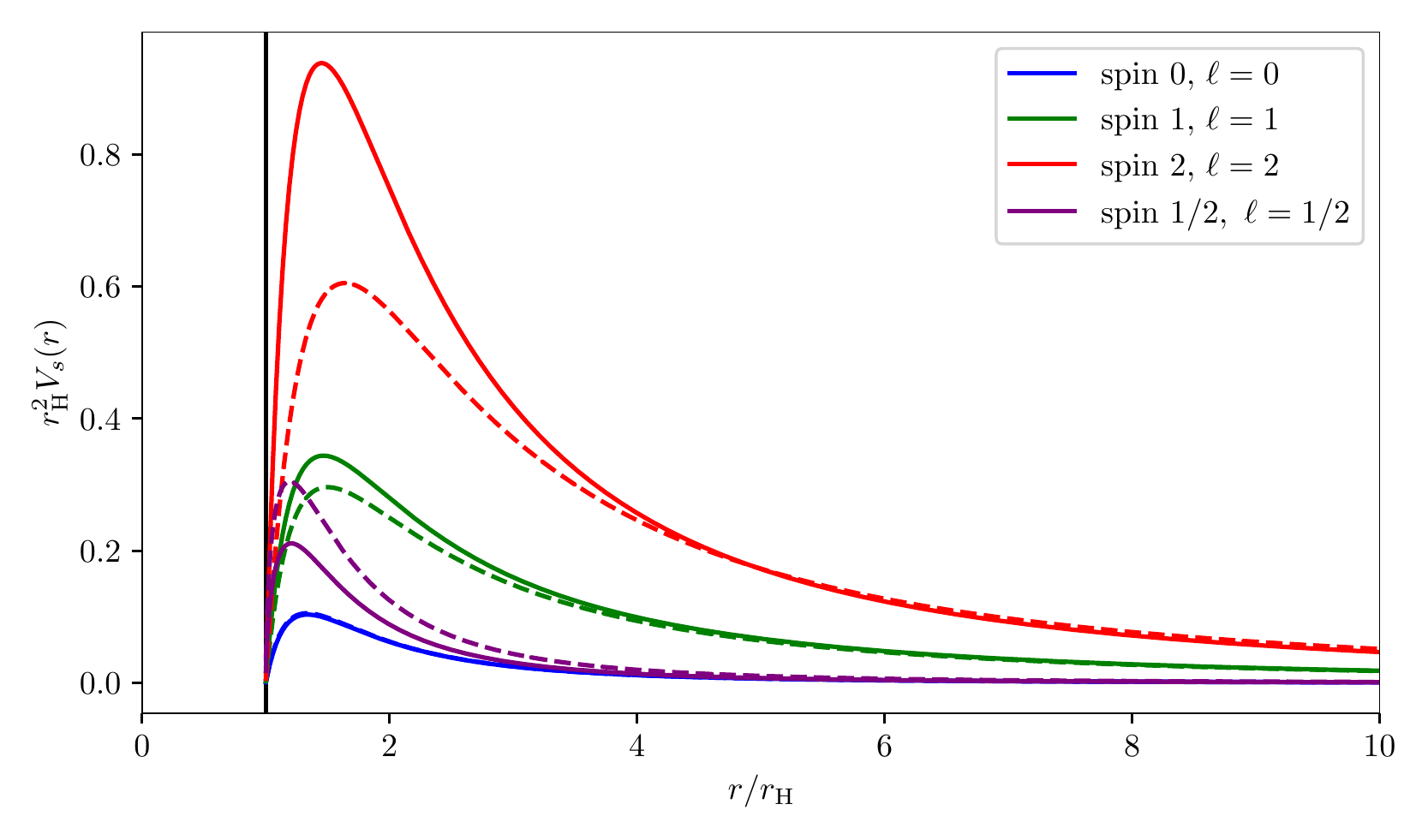}
    \caption{Comparison of the potentials for a polymerized BH with $\epsilon = 0.8$ and $a_0 = 10^{-10}r\mrm{S}^2$ (solid lines) and for the Schwarzschild metric (dashed lines). The vertical black line denotes the BH horizon.}
    \label{fig:LQG_tc}
\end{figure}

In figure \ref{fig:LQG_tc} we show these potentials compared to the Schwarzschild ones for $\epsilon=0.8$ and $a_0=10^{-10}r\mrm{S}^2$. For this particular example, the spin 0 potentials almost coincide because of the cancellation of most of the corrections due to the choice of the angular mode $\ell=0$.

\section*{Conclusion}
\label{sec:conclusion}

In this paper we have studied the dynamics of massless fields of all spins in the general spherically-symmetric and static black hole metrics \eqref{eq:metric}, deriving a generic one-dimensional radial Teukolsky equation. For the spin 1 and spin $1/2$ cases, we have computed the short-ranged potentials, following the transformation developed by Chandrasekhar, thus completing the existing literature on spins 0 and 2. We have applied our general formalism to three examples, namely charged black holes, higher-dimensional black holes and polymerized black holes (coming from effective models of loop quantum gravity), and compared them to the case of Schwarzschild black holes. We have seen that the resulting potentials can largely deviate from the Schwarzschild case, in particular for spin 2 fields. The two main applications of these potentials is the computation of quasi-normal modes and Hawking radiation. The latter will be the subject of the companion paper \cite{AAGLS2}, which will tackle the in-depth computation of Hawking radiation for modified gravity black holes in order to identify critical differences with the emission by a standard Schwarzschild black hole.

\appendix

\section{Details on the radial Teukolsky equations}
\label{Teukolsky details}

In this appendix we give the detailed equations leading to the radial Teukolsky equations for all spins, which take the general form \eqref{eq:teukolsky_general}. We recall that prime denotes a radial derivative.

\paragraph*{\textbf{Massive spin \boldsymbol{$0$}.}}

The massive spin 0 field satisfies the equation of motion \eqref{eq:spin_0_NP}. Using the metric \eqref{eq:metric}, this becomes
\begin{equation}
	-\partial_{t}^2\phi+\f{G}{H}\left(\f{1}{\sin\theta}\partial_\theta(\sin\theta\,\partial_\theta)+\csc^2\theta\,\partial_\varphi^2\right)\phi+\f{\sqrt{FG}}{H}\left(\sqrt{FG}H\phi'\right)'+Gm^2_\phi\phi=0\,.
\end{equation}
Using the ansatz \eqref{eq:anzatz}, the radial part decouples and becomes
\begin{equation}\label{eq:teukolsky_0}
	\left(\omega^2+Gm_\phi^2-\f{G\lambda_\ell^0}{H}\right)\Phi_0+\f{\sqrt{FG}}{H}\left(\sqrt{FG}H\Phi_0'\right)'=0\,.
\end{equation}
This is the radial Teukolsky equation for a massive spin 0 field.

\paragraph*{\textbf{Massless spin \boldsymbol{$1$}.}}

For the massless spin 1 field, equation \eqref{eq:spin_1_NP} takes the explicit form
\begin{align}
	&-\partial_t^2\phi_0+\sqrt{\f{F}{G}}\left(G'-\f{GH'}{H}\right)\partial_t\phi_0+\left(\f{1}{\sin\theta}\partial_\theta(\sin\theta\,\partial_\theta)+\csc^2\theta\,\partial_\varphi^2+\f{2i\cot\theta}{\sin\theta}\partial_\varphi-1-\cot^2\theta\right)\phi_0\nonumber\\
	&+\left(FG''+\f{FGH''}{2H}-\f{FG'^2}{2G}+\f{FGH'^2}{4H^2}+\f{F'G'}{2}+\f{F'GH'}{4H}+\f{7FG'H'}{4H}\right)\phi_0\nonumber\\
	&+\f{1}{H^2}\sqrt{\f{F}{G}}\left(\sqrt{FG}GH^2\phi_0'\right)'=0\,.
\end{align}
Using the ansatz \eqref{eq:anzatz}, the radial part decouples and becomes
\begin{align}\label{eq:teukolsky_1}
	&\left(\omega^2+i\omega\sqrt{\f{F}{G}}\left(\f{GH'}{H}-G'\right)+FG''+\f{FGH''}{2H}-\f{FG'^2}{2G}+\f{FGH'^2}{4H^2}+\f{F'G'}{2}+\f{F'GH'}{4H}+\f{7FG'H'}{4H}-\f{G(\lambda_\ell^1+2)}{H}\right)\Phi_1\nonumber\\
	&+\f{1}{H^2}\sqrt{\f{F}{G}}\left(\sqrt{FG}GH^2\Phi_1'\right)'=0\,.
\end{align}
This is the radial Teukolsky equation for a massless spin 1 field.

\paragraph*{\textbf{Massless spin \boldsymbol{$2$}.}}

For the massless spin 2 field, equation \eqref{eq:spin_2_NP} takes the explicit form
\begin{align}
	&-\partial_t^2\psi_0+2\sqrt{\f{F}{G}}\left(G'-\f{GH'}{H}\right)\partial_t\psi_0+\left(\f{1}{\sin\theta}\partial_\theta(\sin\theta\,\partial_\theta)+\csc^2\theta\,\partial_\varphi^2+\f{4i\cot\theta}{\sin\theta}\partial_\varphi-2-4\cot^2\theta\right)\psi_0\nonumber\\
	&+\left(2FG''-\f{FGH''}{H}-\f{FG'^2}{G}+\f{3FGH'^2}{2H^2}+F'G'-\f{F'GH'}{2H}+\f{9FG'H'}{2H}\right)\psi_0\nonumber\\
	&+\f{1}{GH^3}\sqrt{\f{F}{G}}\left(\sqrt{FG}G^2H^3\psi_0'\right)'=0\,.
\end{align}
Using the ansatz \eqref{eq:anzatz}, the radial part decouples and becomes
\begin{align}\label{eq:teukolsky_2}
	&\left(\omega^2+2i\omega\sqrt{\f{F}{G}}\left(\f{GH'}{H}-G'\right)+2FG''-\f{FGH''}{H}-\f{FG'^2}{G}+\f{3FGH'^2}{2H^2}+F'G'-\f{F'GH'}{2H}+\f{9FG'H'}{2H}-\f{G(\lambda_\ell^2+4)}{H}\right)\Phi_2\nonumber\\
	&+\f{1}{GH^3}\sqrt{\f{F}{G}}\left(\sqrt{FG}G^2H^3\Phi_2'\right)'=0\,.
\end{align}
This is the radial Teukolsky equation for a massless spin 2 field.

\paragraph*{\textbf{Massless spin \boldsymbol{$1/2$}.}}

For the massless spin 1/2 field, equation \eqref{eq:spin_12_NP} takes the explicit form
\begin{align}
	&-\partial_t^2\chi_0+\f{1}{2}\sqrt{\f{F}{G}}\left(G'-\f{GH'}{H}\right)\partial_t\chi_0+\left(\f{1}{\sin\theta}\partial_\theta(\sin\theta\,\partial_\theta)+\csc^2\theta\,\partial_\varphi^2+\f{i\cot\theta}{\sin\theta}\partial_\varphi-\f{1}{2}-\f{1}{4}\cot^2\theta\right)\chi_0\nonumber\\
	&+ \left(\f{FG''}{2}+\f{FGH''}{2H}-\f{FG'^2}{4G}+\f{F'G'}{4}+\f{F'GH'}{4H}+\f{3FG'H'}{4H}\right)\chi_0\nonumber\\
	&+\f{1}{H}\sqrt{\f{F}{H}}\left(\sqrt{FH}GH\chi_0'\right)'=0\,.
\end{align}
Using the ansatz \eqref{eq:anzatz}, the radial part decouples and becomes
\begin{align}\label{eq:teukolsky_12}
	&\left(\omega^2+i\omega\f{1}{2}\sqrt{\f{F}{G}}\left(\f{GH'}{H}-G'\right)+\f{FG''}{2}+\f{FGH''}{2H}-\f{FG'^2}{4G}+\f{3FG'H'}{4H}+\f{F'G'}{4}+\f{F'GH'}{4H}-\f{G(\lambda_\ell^{1/2}+1)}{H}\right)\Phi_{1/2}\nonumber\\
	&+\f{1}{H}\sqrt{\f{F}{H}}\left(\sqrt{FH}GH\Phi_{1/2}'\right)'=0\,.
\end{align}
This is the radial Teukolsky equation for a massless spin 1/2 field.

\paragraph*{\textbf{Massless spin \boldsymbol{$3/2$}.}}

For the massless spin 3/2 field, equation \eqref{eq:spin_32_NP} takes the explicit form
\begin{align}
	&-\partial_t^2H_0+\f{3}{2}\sqrt{\f{F}{G}}\left(G'-\f{GH'}{H}\right)\partial_tH_0+\left(\f{1}{\sin\theta}\partial_\theta(\sin\theta\,\partial_\theta)+\csc^2\theta\,\partial_\varphi^2+\f{3i\cot\theta}{\sin\theta}\partial_\varphi-\f{3}{2}-\f{9}{4}\cot^2\theta\right)H_0\nonumber\\
	&+ \left(\f{3FG''}{2}-\f{3FG'^2}{4G}+\f{3FGH'^2}{4H^2}+\f{3F'G'}{4}+\f{3FG'H'}{H}\right)H_0\nonumber\\
	&+\f{1}{GH^2}\sqrt{\f{F}{H}}\left(\sqrt{FH}G^2H^2H_0'\right)'=0\,.
\end{align}
Using the ansatz \eqref{eq:anzatz}, the radial part decouples and becomes
\begin{align}\label{eq:teukolsky_32}
	&\left(\omega^2+i\omega\f{3}{2}\sqrt{\f{F}{G}}\left(\f{GH'}{H}-G'\right)+\f{3FG''}{2}-\f{3FG'^2}{4G}+\f{3FGH'^2}{4H^2}+\f{3F'G'}{4}+\f{3FG'H'}{H}-\f{G(\lambda_\ell^{3/2}+3)}{H}\right)\Phi_{3/2}\nonumber\\
	&+\f{1}{GH^2}\sqrt{\f{F}{H}}\left(\sqrt{FH}G^2H^2\Phi_{3/2}'\right)'=0\,.
\end{align}
This is the radial Teukolsky equation for a massless spin 3/2 field.

Using the master NP equation \eqref{eq:master_NP} directly, one can obtain a master Teukolsky equation for spins $s>0$, which can be written under the form
\begin{align}
    &-\dfrac{H}{G}\partial_t^2\Phi + s\sqrt{\dfrac{F}{G}}\left( H\dfrac{G^\prime}{G} - H^\prime \right)\partial_t\Phi + FH\partial_r^2\Phi_s + \left( \dfrac{F^\prime H}{2} + (s+1/2)\dfrac{FG^\prime H}{G} + (s+1)FH^\prime \right)\partial_r \Phi_s \nonumber \\
	&+\left( \dfrac{1}{\sin(\theta)}\partial_\theta (\sin(\theta)\partial_\theta) + \dfrac{2is\cot(\theta)}{\sin(\theta)}\partial_\varphi + \dfrac{1}{\sin(\theta)^2}\partial_\varphi^2 - s - s^2\cot(\theta)^2 \right)\Phi_s +\Bigg[ s\dfrac{FG^{\prime\prime}H}{G} \nonumber \\ 
	&  + \dfrac{3s - 2s^2}{2}FH^{\prime\prime} - \dfrac{s}{2}\dfrac{FG^{\prime 2}H}{G^2} + \dfrac{2s^2 - s}{4}\dfrac{FH^{\prime 2}}{H} + \dfrac{s}{2}\dfrac{F^\prime G^\prime H}{G} + \dfrac{3s - 2s^2}{4} F^\prime H^\prime + \dfrac{2s^2 + 5s}{4}\dfrac{FG^\prime H^\prime}{G} \Bigg]\Phi_s = 0 \nonumber \\
	\iff &\sqrt{\dfrac{F}{G}}\dfrac{1}{(GH)^s}\partial_r \left( \sqrt{FG}(GH)^s H\partial_r \Phi_s \right) + \Bigg[ \dfrac{H}{G}\omega^2 + is\omega\sqrt{\dfrac{F}{G}}\left(  H^\prime -  H\dfrac{G^\prime}{G} \right) + s\dfrac{FG^{\prime\prime}H}{G} + \dfrac{s(3 - 2s)}{2}FH^{\prime\prime} \nonumber \\
	&  - \dfrac{s}{2}\dfrac{FG^{\prime 2}H}{G^2} + \dfrac{s(2s - 1)}{4}\dfrac{FH^{\prime 2}}{H} + \dfrac{s}{2}\dfrac{F^\prime G^\prime H}{G} + \dfrac{s(3 - 2s)}{4} F^\prime H^\prime + \dfrac{s(2s + 5)}{4}\dfrac{FG^\prime H^\prime}{G} - \lambda_{l}^s - 2s\Bigg] \Phi_s = 0\,, \label{eq:Teukolsky_master}
\end{align}
where we have used the ansatz \eqref{eq:anzatz} to obtain the final expression. One can observe that this equation is also valid for $s = 0$, as shows a comparison with equation \eqref{eq:teukolsky_0}.

\bibliography{biblio}

\end{document}